\documentclass[runningheads]{llncs}
\usepackage{graphicx}
\usepackage{cite}
\usepackage{amsmath,amssymb,amsfonts}
\usepackage{algorithmic}
\usepackage{microtype}
\usepackage[varqu]{zi4}
\usepackage{verbatim}
\usepackage{graphicx}
\usepackage{textcomp}
\usepackage{xcolor}
\PassOptionsToPackage{hyphens}{url}
\usepackage[hidelinks]{hyperref}
\usepackage{xspace}
\def\BibTeX{{\rm B\kern-.05em{\sc i\kern-.025em b}\kern-.08em
    T\kern-.1667em\lower.7ex\hbox{E}\kern-.125emX}}

\usepackage{mdframed}
\usepackage{soul}

\newcommand{\code}[1]{\texttt{{\small #1}}}

\usepackage{todonotes}

\usepackage{soul}

\usepackage{cleveref}
\crefname{section}{§}{§§}
\Crefname{section}{§}{§§}
\crefformat{section}{\S#2#1#3}
\crefname{figure}{Figure}{Figures}
\crefname{table}{Table}{Tables}

\usepackage{booktabs}

\usepackage{listings}
\usepackage{xcolor}
\usepackage{textcomp} 
\usepackage{tikz}

\definecolor{keyword_color}{rgb}{0.8431372549019608, 0.0, 0.023529411764705882}
\definecolor{identifier_color}{rgb}{0.8980392156862745, 0.36470588235294116, 0.0}
\definecolor{string_color}{rgb}{0.1607843137254902, 0.1843137254901961, 0.40784313725490196}
\definecolor{number_color}{rgb}{0.027450980392156862, 0.396078431372549, 0.796078431372549}
\definecolor{annotation_color}{rgb}{0.43529411764705883, 0.23529411764705882, 0.7686274509803922}
\definecolor{grey}{rgb}{0.6, 0.6, 0.6}

\makeatletter
\newcommand*\idstyle{%
        \expandafter\id@style\the\lst@token\relax
}
\def\id@style#1#2\relax{%
        \ifcat#1\relax\else
                \ifnum`#1=\uccode`#1%
                        \color{number_color}
                \else
                        \color{identifier_color}
                \fi
        \fi
}
\makeatother

\lstdefinelanguage{graphql}{
  identifierstyle=\idstyle,
  delim=[s][\color{string_color}]{"}{"},%
  literate=
     *{0}{{{\color{number_color}0}}}{1}
      {1}{{{\color{number_color}1}}}{1}
      {2}{{{\color{number_color}2}}}{1}
      {3}{{{\color{number_color}3}}}{1}
      {4}{{{\color{number_color}4}}}{1}
      {5}{{{\color{number_color}5}}}{1}
      {6}{{{\color{number_color}6}}}{1}
      {7}{{{\color{number_color}7}}}{1}
      {8}{{{\color{number_color}8}}}{1}
      {9}{{{\color{number_color}9}}}{1}
      {query\ }{{{\color{keyword_color}{query }}}}{1}
      {mutation\ }{{{\color{keyword_color}{mutation }}}}{1}
      {schema}{{{\color{keyword_color}{schema }}}}{1}
      {type}{{{\color{keyword_color}{type }}}}{1}
      {input\ }{{{\color{keyword_color}{input }}}}{1}
      {!}{{{\color{keyword_color}{!}}}}{1}
      {@deprecated}{{{\color{annotation_color}{@deprecated}}}}{1}
      {...\ on}{{{\color{keyword_color}{\texttt{... on }}}}}{1}
      ,
}

\lstnewenvironment{lstgql}%
{\lstset{
  language=graphql,
  basicstyle=\fontsize{6.5}{8}\ttfamily,
  columns=fullflexible,
  numbers=left,
  numberstyle=\fontsize{6.5}{8}\ttfamily\color{grey},
  xleftmargin=3em,
  belowskip=0em,
  aboveskip=0em,
}}
{}

\lstnewenvironment{lstgqlsmall}%
{\lstset{
  language=graphql,
  basicstyle=\scriptsize\ttfamily,
  columns=fullflexible,
  numbers=none,
  xleftmargin=3em,
}}
{}

\colorlet{punct}{red!60!black}
\definecolor{background}{HTML}{EEEEEE}
\definecolor{delim}{RGB}{20,105,176}
\colorlet{numb}{magenta!60!black}
\lstdefinelanguage{json}{
    basicstyle=\normalfont\ttfamily,
    numbers=left,
    numberstyle=\scriptsize,
    stepnumber=1,
    numbersep=8pt,
    showstringspaces=false,
    breaklines=true,
    frame=lines,
    backgroundcolor=\color{background},
    literate=
     *{0}{{{\color{numb}0}}}{1}
      {1}{{{\color{numb}1}}}{1}
      {2}{{{\color{numb}2}}}{1}
      {3}{{{\color{numb}3}}}{1}
      {4}{{{\color{numb}4}}}{1}
      {5}{{{\color{numb}5}}}{1}
      {6}{{{\color{numb}6}}}{1}
      {7}{{{\color{numb}7}}}{1}
      {8}{{{\color{numb}8}}}{1}
      {9}{{{\color{numb}9}}}{1}
      {:}{{{\color{punct}{:}}}}{1}
      {,}{{{\color{punct}{,}}}}{1}
      {\{}{{{\color{delim}{\{}}}}{1}
      {\}}{{{\color{delim}{\}}}}}{1}
      {[}{{{\color{delim}{[}}}}{1}
      {]}{{{\color{delim}{]}}}}{1},
}

\newcommand{\CommercialNumRawFiles}{33\xspace}

\newcommand{\NumIncompleteSchemas}{30,017\xspace}
\newcommand{\NumMergedSchemasWithQueryType}{5,603\xspace}
\newcommand{\NumRecoveredSchemas}{2,453\xspace}
\newcommand{\NumRecoveryRate}{43.8\%\xspace}

\newcommand{\Schemas}{8,399\xspace}
\newcommand{\MergedSchemas}{1,127\xspace}
\newcommand{\WithMutations}{5,699\xspace}
\newcommand{\WithMutationsPercent}{67.9\%\xspace}
\newcommand{\WithSubscriptions}{2,096\xspace}
\newcommand{\WithSubscriptionsPercent}{25.0\%\xspace}
\newcommand{\MedianOTs}{6\xspace}
\newcommand{\MedianFieldsOTs}{3\xspace}
\newcommand{\MedianInputOTs}{6\xspace}
\newcommand{\MedianFieldsInputOTs}{3\xspace}
\newcommand{\WithInterfaces}{2,377\xspace}
\newcommand{\WithInterfacesPercent}{28.3\%\xspace}
\newcommand{\WithUnions}{506\xspace}
\newcommand{\WithUnionsPercent}{6.0\%\xspace}
\newcommand{\WithDirectives}{160\xspace}
\newcommand{\WithDirectivesPercent}{1.9\%\xspace}

\newcommand{\MutationFieldNamesSchemas}{49.3\%\xspace}

\newcommand{\CamelFieldSchemas}{53.9\%\xspace}
\newcommand{\SnakeField}{30.6\%\xspace}
\newcommand{\SnakeFieldSchemas}{0.5\%\xspace}

\newcommand{\PascalTypeSchemas}{91.8\%\xspace}

\newcommand{\PascalEnumSchemas}{96.8\%\xspace}

\newcommand{\AllCapsEnumSchemas}{35.7\%\xspace}

\newcommand{\InputPostfixSchemas}{71.6\%\xspace}
\newcommand{\WorstExponential}{3,219\xspace}
\newcommand{\WorstExponentialPercent}{38.3\%\xspace}
\newcommand{\WorstZero}{1,048\xspace}
\newcommand{\WorstZeroPercent}{12.5\%\xspace}
\newcommand{\WorstOne}{3,112\xspace}
\newcommand{\WorstOnePercent}{37.1\%\xspace}
\newcommand{\WorstTwo}{785\xspace}
\newcommand{\WorstTwoPercent}{9.3\%\xspace}
\newcommand{\WorstThree}{186\xspace}
\newcommand{\WorstThreePercent}{2.2\%\xspace}
\newcommand{\WorstFour}{34\xspace}
\newcommand{\WorstFourPercent}{0.4\%\xspace}
\newcommand{\WorstFive}{9\xspace}
\newcommand{\WorstFivePercent}{0.1\%\xspace}
\newcommand{\WorstSix}{6\xspace}
\newcommand{\WorstSixPercent}{0.1\%\xspace}
\newcommand{\NumSchemasLists}{7,351\xspace}
\newcommand{\NumSchemasListsPercent}{87.5\%\xspace}

\newcommand{\NumSchemasNoSlicing}{5,335\xspace}
\newcommand{\NumSchemasNoSlicingPercent}{63.5\%\xspace}
\newcommand{\NumSchemasMixedSlicing}{1,771\xspace}
\newcommand{\NumSchemasMixedSlicingPercent}{21.1\%\xspace}
\newcommand{\NumSchemasPureSlicing}{245\xspace}
\newcommand{\NumSchemasPureSlicingPercent}{2.9\%\xspace}
\newcommand{\NumSchemasConnections}{2,073\xspace}
\newcommand{\NumSchemasConnectionsPercent}{24.7\%\xspace}
\newcommand{\NumSchemasConnectionsNoSlicing}{1,397\xspace}
\newcommand{\NumSchemasConnectionsNoSlicingPercent}{16.6\%\xspace}
\newcommand{\NumSchemasConnectionsMixedSlicing}{48\xspace}
\newcommand{\NumSchemasConnectionsMixedSlicingPercent}{0.6\%\xspace}
\newcommand{\NumSchemasConnectionsPureSlicing}{628\xspace}
\newcommand{\NumSchemasConnectionsPureSlicingPercent}{7.5\%\xspace}
\newcommand{\QualitySchemas}{1,739\xspace}
\newcommand{\QualityMergedSchemas}{10\xspace}
\newcommand{\QualityWithMutations}{1,672\xspace}
\newcommand{\QualityWithMutationsPercent}{96.1\%\xspace}
\newcommand{\QualityWithSubscriptions}{1,113\xspace}
\newcommand{\QualityWithSubscriptionsPercent}{64.0\%\xspace}
\newcommand{\QualityMedianOTs}{35\xspace}
\newcommand{\QualityMedianFieldsOTs}{3\xspace}
\newcommand{\QualityMedianInputOTs}{43\xspace}
\newcommand{\QualityMedianFieldsInputOTs}{3\xspace}
\newcommand{\QualityWithInterfaces}{1,395\xspace}
\newcommand{\QualityWithInterfacesPercent}{80.2\%\xspace}
\newcommand{\QualityWithUnions}{330\xspace}
\newcommand{\QualityWithUnionsPercent}{19.0\%\xspace}
\newcommand{\QualityWithDirectives}{26\xspace}
\newcommand{\QualityWithDirectivesPercent}{1.5\%\xspace}

\newcommand{\QualityMutationFieldNamesSchemas}{62.9\%\xspace}

\newcommand{\QualityCamelFieldSchemas}{8.2\%\xspace}

\newcommand{\QualitySnakeFieldSchemas}{0.1\%\xspace}

\newcommand{\QualityPascalTypeSchemas}{82.1\%\xspace}

\newcommand{\QualityPascalEnumSchemas}{96.4\%\xspace}

\newcommand{\QualityAllCapsEnumSchemas}{12.1\%\xspace}

\newcommand{\QualityInputPostfixSchemas}{68.2\%\xspace}
\newcommand{\QualityWorstExponential}{1,414\xspace}
\newcommand{\QualityWorstExponentialPercent}{81.3\%\xspace}
\newcommand{\QualityWorstZero}{0\xspace}
\newcommand{\QualityWorstZeroPercent}{0.0\%\xspace}
\newcommand{\QualityWorstOne}{182\xspace}
\newcommand{\QualityWorstOnePercent}{10.5\%\xspace}
\newcommand{\QualityWorstTwo}{88\xspace}
\newcommand{\QualityWorstTwoPercent}{5.1\%\xspace}
\newcommand{\QualityWorstThree}{40\xspace}
\newcommand{\QualityWorstThreePercent}{2.3\%\xspace}
\newcommand{\QualityWorstFour}{7\xspace}
\newcommand{\QualityWorstFourPercent}{0.4\%\xspace}
\newcommand{\QualityWorstFive}{4\xspace}
\newcommand{\QualityWorstFivePercent}{0.2\%\xspace}
\newcommand{\QualityWorstSix}{4\xspace}
\newcommand{\QualityWorstSixPercent}{0.2\%\xspace}
\newcommand{\QualityNumSchemasLists}{1,739\xspace}
\newcommand{\QualityNumSchemasListsPercent}{100.0\%\xspace}

\newcommand{\QualityNumSchemasNoSlicing}{385\xspace}
\newcommand{\QualityNumSchemasNoSlicingPercent}{22.1\%\xspace}
\newcommand{\QualityNumSchemasMixedSlicing}{1,265\xspace}
\newcommand{\QualityNumSchemasMixedSlicingPercent}{72.7\%\xspace}
\newcommand{\QualityNumSchemasPureSlicing}{89\xspace}
\newcommand{\QualityNumSchemasPureSlicingPercent}{5.1\%\xspace}
\newcommand{\QualityNumSchemasConnections}{1,365\xspace}
\newcommand{\QualityNumSchemasConnectionsPercent}{78.5\%\xspace}
\newcommand{\QualityNumSchemasConnectionsNoSlicing}{1,073\xspace}
\newcommand{\QualityNumSchemasConnectionsNoSlicingPercent}{61.7\%\xspace}
\newcommand{\QualityNumSchemasConnectionsMixedSlicing}{31\xspace}
\newcommand{\QualityNumSchemasConnectionsMixedSlicingPercent}{1.8\%\xspace}
\newcommand{\QualityNumSchemasConnectionsPureSlicing}{261\xspace}
\newcommand{\QualityNumSchemasConnectionsPureSlicingPercent}{15.0\%\xspace}
\newcommand{\CommercialSchemas}{16\xspace}

\newcommand{\CommercialWithMutations}{11\xspace}
\newcommand{\CommercialWithMutationsPercent}{68.8\%\xspace}
\newcommand{\CommercialWithSubscriptions}{0\xspace}
\newcommand{\CommercialWithSubscriptionsPercent}{0.0\%\xspace}
\newcommand{\CommercialMedianOTs}{60\xspace}
\newcommand{\CommercialMedianFieldsOTs}{3\xspace}
\newcommand{\CommercialMedianInputOTs}{44\xspace}
\newcommand{\CommercialMedianFieldsInputOTs}{2\xspace}
\newcommand{\CommercialWithInterfaces}{11\xspace}
\newcommand{\CommercialWithInterfacesPercent}{68.8\%\xspace}
\newcommand{\CommercialWithUnions}{8\xspace}
\newcommand{\CommercialWithUnionsPercent}{50.0\%\xspace}
\newcommand{\CommercialWithDirectives}{2\xspace}
\newcommand{\CommercialWithDirectivesPercent}{12.5\%\xspace}

\newcommand{\CommercialMutationFieldNamesSchemas}{9.1\%\xspace}

\newcommand{\CommercialCamelFieldSchemas}{12.5\%\xspace}

\newcommand{\CommercialSnakeFieldSchemas}{0.0\%\xspace}

\newcommand{\CommercialPascalTypeSchemas}{62.5\%\xspace}

\newcommand{\CommercialPascalEnumSchemas}{81.3\%\xspace}

\newcommand{\CommercialAllCapsEnumSchemas}{56.3\%\xspace}

\newcommand{\CommercialInputPostfixSchemas}{23.1\%\xspace}
\newcommand{\CommercialWorstExponential}{14\xspace}
\newcommand{\CommercialWorstExponentialPercent}{87.5\%\xspace}
\newcommand{\CommercialWorstZero}{0\xspace}
\newcommand{\CommercialWorstZeroPercent}{0.0\%\xspace}
\newcommand{\CommercialWorstOne}{0\xspace}
\newcommand{\CommercialWorstOnePercent}{0.0\%\xspace}
\newcommand{\CommercialWorstTwo}{1\xspace}
\newcommand{\CommercialWorstTwoPercent}{6.3\%\xspace}
\newcommand{\CommercialWorstThree}{1\xspace}
\newcommand{\CommercialWorstThreePercent}{6.3\%\xspace}
\newcommand{\CommercialWorstFour}{0\xspace}
\newcommand{\CommercialWorstFourPercent}{0.0\%\xspace}
\newcommand{\CommercialWorstFive}{0\xspace}
\newcommand{\CommercialWorstFivePercent}{0.0\%\xspace}
\newcommand{\CommercialWorstSix}{0\xspace}
\newcommand{\CommercialWorstSixPercent}{0.0\%\xspace}
\newcommand{\CommercialNumSchemasLists}{16\xspace}
\newcommand{\CommercialNumSchemasListsPercent}{100.0\%\xspace}

\newcommand{\CommercialNumSchemasNoSlicing}{10\xspace}
\newcommand{\CommercialNumSchemasNoSlicingPercent}{62.5\%\xspace}
\newcommand{\CommercialNumSchemasMixedSlicing}{6\xspace}
\newcommand{\CommercialNumSchemasMixedSlicingPercent}{37.5\%\xspace}
\newcommand{\CommercialNumSchemasPureSlicing}{0\xspace}
\newcommand{\CommercialNumSchemasPureSlicingPercent}{0.0\%\xspace}
\newcommand{\CommercialNumSchemasConnections}{9\xspace}
\newcommand{\CommercialNumSchemasConnectionsPercent}{56.3\%\xspace}
\newcommand{\CommercialNumSchemasConnectionsNoSlicing}{1\xspace}
\newcommand{\CommercialNumSchemasConnectionsNoSlicingPercent}{6.3\%\xspace}
\newcommand{\CommercialNumSchemasConnectionsMixedSlicing}{2\xspace}
\newcommand{\CommercialNumSchemasConnectionsMixedSlicingPercent}{12.5\%\xspace}
\newcommand{\CommercialNumSchemasConnectionsPureSlicing}{6\xspace}
\newcommand{\CommercialNumSchemasConnectionsPureSlicingPercent}{37.5\%\xspace}

\newcommand{\QualityCorpusPercentOfGithubCorpus}{20.7\%\xspace}
\newcommand{\SnakeCasePercentOfNonCamelCase}{90.3\%\xspace}
\newcommand{\SomeSnakeCasePercent}{37.3\%\xspace}
\newcommand{\SuperLinear}{50.5\%\xspace}
\newcommand{\QualitySuperLinear}{89.5\%\xspace}
\newcommand{\CommercialSuperLinear}{100.0\%\xspace}

\begin{document}
\title{An Empirical Study of GraphQL Schemas}
\author{
  Erik Wittern\inst{1}
\and
  Alan Cha\inst{1}
\and
  James C. Davis\inst{2}
\and\\
  Guillaume Baudart\inst{1}
\and
  Louis Mandel\inst{1}
}
\authorrunning{E.Wittern et al.}
%
\institute{
  IBM Research, USA \\
  \email{\{witternj,lmandel\}@us.ibm.com, \{alan.cha1,guillaume.baudart\}@ibm.com}
\and
  Virginia Tech, USA
  \email{davisjam@vt.edu}
}
\maketitle              
\begin{abstract}
GraphQL is a query language for APIs and a runtime to execute queries. Using GraphQL queries, clients define precisely what data they wish to retrieve or mutate on a server, leading to fewer round trips and reduced response sizes.
Although interest in GraphQL is on the rise, with increasing adoption at major organizations, little is known about what GraphQL interfaces look like in practice.
This lack of knowledge
makes it hard for providers to understand what practices promote idiomatic, easy-to-use APIs, and what pitfalls to avoid.

To address this gap, we study the design of GraphQL interfaces in practice by analyzing their schemas -- the descriptions of their exposed data types and the possible operations on the underlying data.
We base our study on two novel corpuses of GraphQL schemas, one of \CommercialSchemas commercial GraphQL schemas and the other of \Schemas GraphQL schemas mined from GitHub projects. We make available to other researchers those schemas mined from GitHub whose licenses permit redistribution. We also make available the scripts to mine the whole corpus.
Using the two corpuses, we
  characterize the size of schemas
    and
  their use of GraphQL features
and
  assess the use of both prescribed and organic naming conventions.
We also report that a majority of APIs are susceptible to denial of service through complex queries, posing real security risks previously discussed only in theory.
We also assess ways in which GraphQL APIs attempt to address these concerns.

\keywords{GraphQL  \and Web APIs \and Practices.}
\end{abstract}
\section{Introduction}
GraphQL is a \emph{query language} for web APIs, and a corresponding \emph{runtime} for executing queries.
To offer a GraphQL API, providers define a \emph{schema} containing the available data \emph{types}, their \emph{relations}, and the possible \emph{operations} on that data.
Clients send \emph{queries} that precisely define the data they wish to retrieve or mutate. The server implementing the GraphQL API executes the query, and returns exactly the requested data.
\cref{fig:example-query} shows, on the left, an example query for GitHub's GraphQL API~\cite{GitHubGraphQL:2019}. It aims to retrieve the \code{description} of the \code{graphql-js} repository owned by \code{graphql}. The query, in case that this owner is an \code{Organization}, further requests the \code{totalCount} of all members of that organization, and the \code{name}s of the first two of them. The right hand side of \cref{fig:example-query} shows the response produced by GitHub's GraphQL API after executing that query,\footnote{We anonymized the returned names.} which contains exactly the requested data.

\begin{figure}[h]
  \centering
  \begin{minipage}{0.54\textwidth}
  \begin{lstgql}
query {
  repository(name: "graphql-js", owner: "graphql") {
    description
    owner {
      ... on Organization {
        membersWithRole(first: 2) {
          totalCount
          nodes {
            name
          }
} } } } }
  \end{lstgql}
  \end{minipage}
  \vline
  \begin{minipage}{0.45\textwidth}
    \begin{lstgql}
{ "data": {
  "repository": {
    "description": "A reference imple...",
    "owner": {
      "membersWithRole": {
        "totalCount": 4,
        "nodes": [
          {"name": "Member 1"},
          {"name": "Member 2"}
        ]
} } } } }
    \end{lstgql}
    \end{minipage}
  \caption{Example of a GraphQL query (left), and corresponding JSON response (right).}
  \label{fig:example-query}
\end{figure}

GraphQL is seeing adoption at major organizations thanks in part to its advantages for performance and usability.
In some use-cases, allowing users to precisely state data requirements using GraphQL queries can lead to fewer request-response roundtrips and smaller response sizes as compared to other API paradigms, e.g., REST-like APIs~\cite{Brito:2019}. GraphQL prescribes a statically \emph{typed} interface, which drives developer tooling like GraphiQL, an online IDE helping developers explore schemas and write and validate queries~\cite{GraphiQL}, or type-based data mocking for testing services\cite{GraphQL-Faker}.
Major organizations have begun to embrace it, including
GitHub~\cite{GitHubGraphQL:2019},
Yelp~\cite{YelpGraphQL:2017},
The New York Times~\cite{NYTimesGraphQL:2017},
or Shopify~\cite{ShopifyGraphQL:2019}.

As any new technology is deployed, users begin to follow useful patterns and identify best practices and anti-patterns.
Our aim is to shed light on emerging GraphQL uses and practices, in the spirit of similar studies for REST(-like) APIs~\cite{Palma:2015,Petrillo:2016,Gamez:2017}.
By studying technological practices in the GraphQL context, we benefit the entire GraphQL community:
Our study will help GraphQL providers build idiomatic, easy-to-use GraphQL APIs, and avoid pitfalls others have experienced before.
Our findings also inform tool developers about the practices that are more (and less) important to support.
Obviously GraphQL consumers will benefit from the resulting well-designed GraphQL APIs and effective tool support. 
And finally, our contributions may influence the evolution of GraphQL itself, as we highlight challenges that the specification may eventually address.


Specifically, the contributions of this work are:

\begin{itemize}
 \item We present two novel GraphQL schema corpuses, derived respectively from commercial GraphQL deployments and open-source projects (\cref{sec:corpuses}). We make parts of the open-source corpus -- as permitted by schema licenses -- publicly available for other researchers~\cite{Wittern:2019}, and also share the scripts to reproduce the whole open-source corpus~\cite{Wittern:2019b}.
 \item We analyze our corpuses for common schema characteristics, naming conventions, and worst-case response sizes, and describe practices that address large responses (\cref{sec:analysis}).
\end{itemize}

In brief, we find that:
(1) There are significant differences between commercial and open-source schemas;
(2) Schemas commonly follow naming conventions, both documented and not;
(3) A majority of schemas have large worst-case response sizes, which schema developers and endpoint providers should consider; and
(4) Mechanisms to avoid these large response sizes are applied inconsistently.

\section{Background} \label{sec:background}

As sketched above, a schema describes the types of data offered by a GraphQL API, the relations between those types, and possible operations on them.
In this section, we outline selected concepts related to GraphQL schemas.
GraphQL providers can define schemas either programmatically using libraries like graphql-js~\cite{GraphQL-js}, or they can define them declaratively using the \emph{Schema Definition Language} (SDL).
\cref{fig:schema-example} shows an example schema defined in the SDL.

\begin{figure}[h]
  \centering
  \begin{minipage}{0.57\textwidth}
    \begin{lstgql}
schema {
  query: Query
  mutation: Mutation
}
type Mutation {
  createOffice(input: OfficeInput!): Office
}
type Query {
  company(id: ID!): Company
}
type Company {
  id: ID!
  name: String
  address: String
  age: Int @deprecated(reason: "No longer relevant.")
  offices(limit: Int!, after: ID): OfficeConnection
}
    \end{lstgql}
    \end{minipage}
    \vline
    \begin{minipage}{0.3\textwidth}
      \begin{lstgql}
type OfficeConnection {
  totalCount: Int
  nodes: [Office]
  edges: [OfficeEdge]
}
type OfficeEdge {
  node: Office
  cursor: ID
}
type Office {
  id: ID!
  name: String
}
input OfficeInput {
  name: String!
}
      \end{lstgql}
      \end{minipage}
  \caption{Example of a GraphQL schema in the Schema Definition Language (SDL).}
  \label{fig:schema-example}
\end{figure}

The \code{schema} defines \emph{fields} \code{query} and \code{mutation}, one of which forms the entry for any valid query. Every GraphQL schema must contain a \code{Query} operation type, which in this case is the \code{Query} object type. According to this schema, queries can retrieve a \code{company} field that returns a \code{Company} identified by an \code{id} \emph{argument} of type \code{ID}~(the character~``\code{!}'' indicates that the argument is required). The returned \code{Company} again allows queries to retrieve its \code{id}, \code{name}, \code{address}, \code{age}, and/or \code{offices}. The latter requires the user to \code{limit} the number of offices returned. Offices, implementing the \emph{connections pattern} for pagination~\cite{GraphQLDocs-Pagination}, are related to a company via an \code{OfficeConnection}, that contains information about the \code{totalCount} of offices of that company, and grants access to them directly via the \code{nodes} field or indirectly via the \code{edges} field. Querying for an \code{OfficeEdge} allows users to obtain a \code{cursor} that they can use (in subsequent queries) to precisely slice which \code{offices} to retrieve from a \code{Company} via the \code{after} argument.

\begin{lstgqlsmall}
query { company(id: "n3...") { offices(limit: 10, after: "mY...") { edges: { 
      cursor 
      node { name } 
} } } }
\end{lstgqlsmall}
  
The schema further allows to mutate data via the \code{createOffice} field. The data about the office to create is defined in a dedicated input object type called \code{OfficeInput} and passed as an argument. A corresponding query may look like:

\begin{lstgqlsmall}
mutation { createOffice(input: { name: "A new office" }) {
    id 
} }
\end{lstgqlsmall}

In GraphQL, basic types like \code{String}, \code{Int}, or \code{Boolean} are called \emph{scalars}, sets of predefined strings are called \emph{enums}, and complex types that contain fields are called \emph{object types} (e.g., \code{Company} in \cref{fig:schema-example}). GraphQL further allows developers to define \emph{interfaces} that can be implemented by object types or extended by other interfaces, and \emph{unions} which state that data can be of one of multiple object types. For example, in line 5 of \cref{fig:example-query}, \code{... on Organization} is a \emph{type condition} that queries fields on the interface \code{RepositoryOwner} returned by field \code{owner} only if the owner happens to be an \code{Organization}.
Beyond queries that retrieve data, GraphQL schemas may also define a root \code{Mutation} operation type, whose fields define possible mutations, e.g., to create, edit, or delete data.
Input for mutations is defined using arguments, which are (lists of) scalars, enums, or \emph{input} object types.
Finally, GraphQL schemas may contain \emph{directives} that define metadata or behavioral changes associated with field, type, argument or even the whole schema definitions. For example, in \cref{fig:schema-example} the field \code{age} in type \code{Company} is marked with a directive as deprecated. This information could, for example, be displayed by documentation tooling.
Tools or clients can send an \emph{introspection} query to retrieve the latest schema from a GraphQL API.

After defining their GraphQL schema, to offer a GraphQL API a provider must build a mapping between the data types defined in the schema and their representation in the back-end storage system(s).
The provider does this by implementing a \emph{resolver function} for each field defined in the schema, which can retrieve or mutate the corresponding data.
Resolver functions can, for example, interact with databases, other APIs, or dynamically compute a result --- GraphQL is agnostic to their implementation.
To execute a query, a GraphQL runtime validates it against the schema, and then in sequence calls all resolver functions required to fulfill the query.

Although a schema definition does not tell us everything about a GraphQL API (e.g., how its resolver functions are implemented), GraphQL schemas can still tell us about GraphQL practices.
For example, from a GraphQL schema we can learn
  the characteristics of the corresponding GraphQL API,
  the nature of possible queries to its API,
  and the conventions followed in designing it.
Schema definitions thus comprise useful research artifacts.
In the next section we discuss the schema definitions we sampled to understand these and other topics.


\section{Data: Two Novel GraphQL Schema Corpuses} \label{sec:corpuses}
\begin{figure}[h]
  \centering
  \includegraphics[width=0.4\columnwidth]{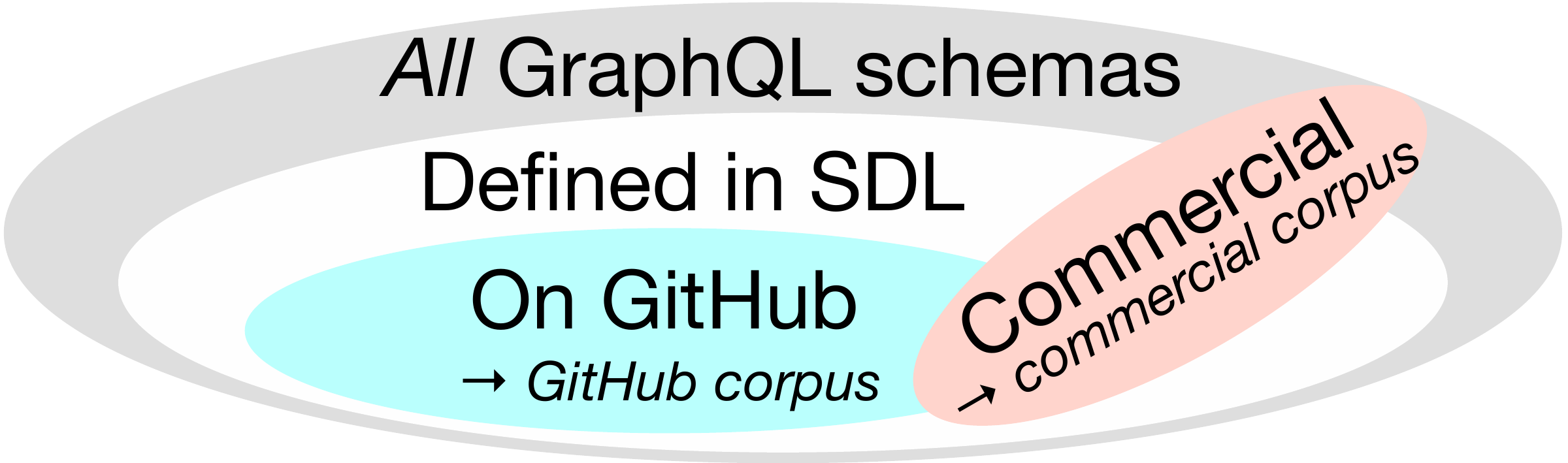}
  \caption{Schema corpuses used in this work. A subset of all GraphQL schemas is defined using the SDL rather than programmatically. We mine the subset hosted on GitHub. Schemas in our commercial corpus can be defined either way, and may be hosted (privately) on GitHub.}
  \label{fig:schema-sets}
\end{figure}

We created two corpuses of GraphQL schemas: one from introspecting publicly accessible commercial GraphQL APIs (\cref{section:corpuses-commercial}), and the other from mining GitHub for GraphQL schema definitions (\cref{section:corpuses-github}).
\cref{fig:schema-sets} illustrates the GraphQL schema populations we sampled to create these corpuses.

In both corpuses, we included only schemas that are parsable (e.g., written in valid SDL syntax) and complete (e.g., contains a \textit{query operation} and definitions of all referenced types). We checked these constraints using the \emph{parsing} and \emph{validation} capabilities offered by the graphql-js reference implementation~\cite{GraphQL-js}, thus ensuring that schemas can be processed and analyzed without risking runtime errors.
We make available the schemas in the open-source corpus -- considering the constraints for redistributing them defined in their licenses~\cite{Wittern:2019}.
We also make available the scripts to collect the whole open-source corpus~\cite{Wittern:2019b}. These scripts contain the schema reconstruction logic described in~\cref{section:corpuses-github}.

\subsection{Commercial Corpus (Schemas Deployed in Practice)} \label{section:corpuses-commercial}
Our commercial corpus (\CommercialSchemas schemas) represents GraphQL schemas written and maintained by professional software developers for business-critical purposes.
This corpus allows us to reason about GraphQL practices in industry.

To identify commercial GraphQL APIs, we started with the community-maintained list provided by \emph{APIs.guru}~\cite{APIsGuru-GraphQLAPIs}.\footnote{We submitted a pull request adding several public GraphQL APIs that were missing from the \emph{APIs.guru} list, but that we found using web searches. The \emph{APIs.guru} maintainers accepted the pull request and we included those schemas in this analysis.}
We manually assessed the documentation for all \CommercialNumRawFiles of the ``Official APIs'' listed on May $1^{st}$ 2019 to remove demo interfaces and non-commercial APIs.
We then used introspection to collect these commercial GraphQL schemas.
After discarding invalid schemas (validity is defined in \cref{section:corpuses-github}), we obtained our final corpus of \CommercialSchemas valid, unique GraphQL schemas maintained by commercial websites.
The corpus includes, among others, schemas of prominent GraphQL APIs like GitHub, Shopify, Yelp, and BrainTree.

\subsection{Open-Source Corpus (Schemas in GitHub Projects)} \label{section:corpuses-github}

\begin{figure}[h!]
  \centering
  \includegraphics[width=0.8\columnwidth]{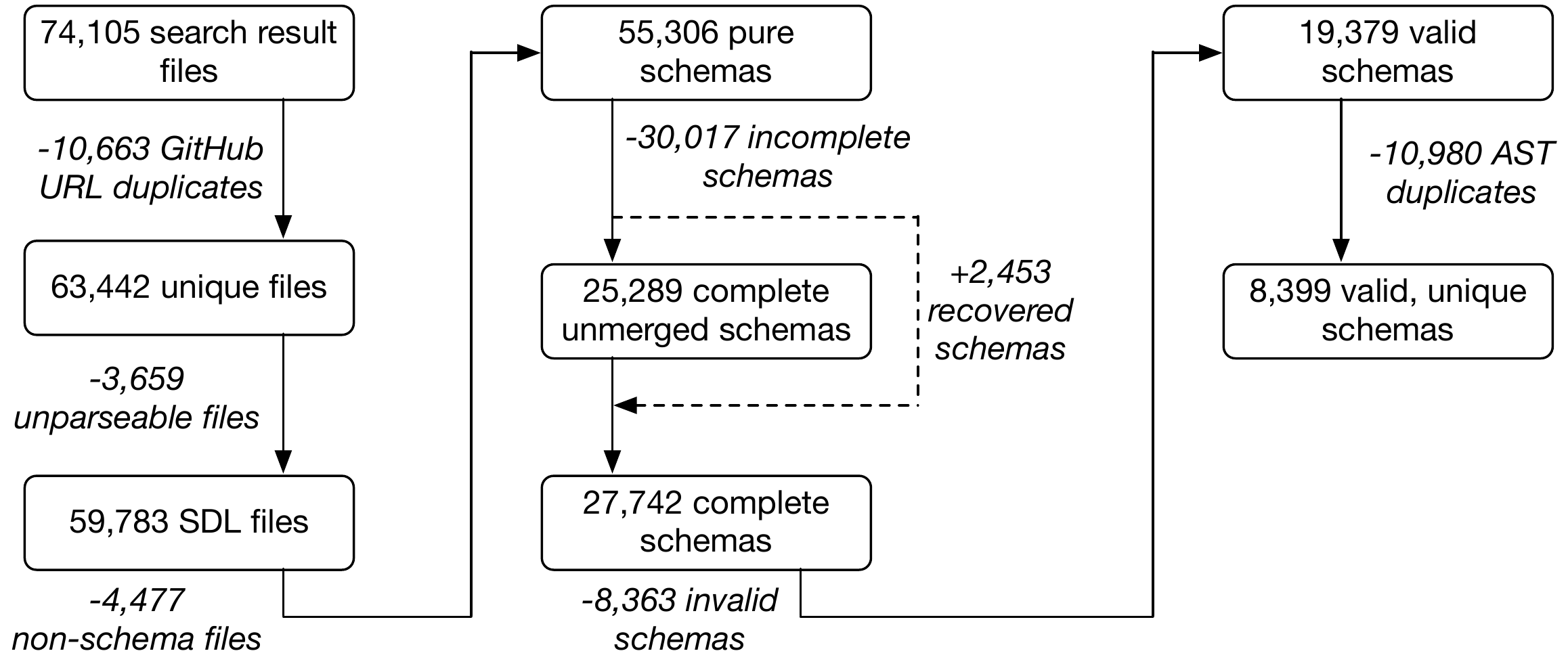}
  \vspace{-10pt}
  \caption{Filters to construct the open-source schema corpus.}
  \label{fig:OpenSourceCorpusFilters}
\end{figure}

Our open-source corpus (\Schemas schemas) provides another perspective on GraphQL practices, depicting (ongoing) development efforts and privately-deployed APIs.
For this corpus, we aimed to collect schema definitions written in the SDL (cf.~\cref{sec:background}) and stored in a GitHub project.
\cref{fig:OpenSourceCorpusFilters} summarizes the stages of our data-collection methodology.

We used GitHub's code search API to obtain \textbf{search result files} that likely contain schemas on May $21^{st}$ 2019 using this query:

\begin{verbatim}
type extension:graphql extension:gql
  size:<min>..<max> fork:false
\end{verbatim}

The pieces of this query have the following meaning.
The \textit{search term \code{type}} is used in any non-trivial schema in the GraphQL SDL.
The \textit{extensions \code{.graphql} and \code{.gql}} are common file suffixes for the GraphQL SDL.
The \textit{file sizes \code{<min>} and \code{<max>}} partitioned the search space by file size in order to work around GitHub's limit on code search query results.
We omitted project forks to avoid duplicates.

We removed duplicates by URL to obtain \textbf{unique files}.
We filtered unparsable files per the graphql-js~\cite{GraphQL-js} reference implementation
to obtain \textbf{SDL files}.
The GraphQL SDL can describe not only schemas, but also executables like queries, mutations, or subscriptions (e.g., the query in \cref{fig:example-query}) or a mixture of both.
Because we are only interested in schemas, we obtained \textbf{pure schemas} by removing any files that contain executables, a.k.a. \textit{executable definitions}~\cite{GraphQLDocs-Executable}.

These steps left us with parsable SDL files, but not all are \textbf{complete}.\footnote{A \textit{complete} schema (1) contains a \textit{query operation} (a \emph{SchemaDefinition} node~\cite{GraphQLDocs-Schema} or a \emph{Query} object type~\cite{GraphQLSchemaLanguage:2018}), and (2) defines all referenced types and directives.} We observed that some schemas contain reference errors, e.g., because they are divided across multiple files for encapsulation.
Supposing that a repository's complete schema(s) can be produced through some combination of its GraphQL files, we used heuristics to try to reconstruct these \textit{partitioned schemas}, thus adding \textbf{recovered schemas} back to our data. 
For every schema that contains a \textit{query operation} but also reference errors, we searched for the missing definitions in the repository's other GraphQL files.
When we found a missing type in another file, we appended that file's contents to the current schema.\footnote{If multiple possible definitions were found, we broke ties under the assumption that developers will use the directory hierarchy to place related files close to each other.}
We repeated this process until we obtained either a complete schema or an unresolvable reference.
Of the \NumIncompleteSchemas \textbf{incomplete schemas}, there are \NumMergedSchemasWithQueryType that contain an \textit{query operation}, meaning they can form the basis of a merged schema, and from these schemas, we were able to recover \NumRecoveredSchemas schemas (\NumRecoveryRate success rate).
This success rate suggests that \textit{distributing GraphQL schema definitions across multiple files is a relatively common practice}.

We obtained \textbf{valid schemas} by removing ones that could not be validated by the graphql-js reference implementation, and finally \textbf{valid, unique schemas} by removing duplicates by testing for abstract syntax tree~(AST) equivalence.
We discarded about half of the remaining schemas during deduplication (\cref{fig:OpenSourceCorpusFilters}).
Inspection suggests many of these schemas were duplicated from examples.

Our final open-source schema corpus contains \Schemas valid, unique GraphQL Schema Definition files, \MergedSchemas of which were recovered through merging.\footnote{We collected data in November 2018 using the same methodology, and found $5,345$ unique schemas, $701$ of which resulted from merging. This reflects a growth of $57\%$ in half a year.}
Although all of these schemas are valid, some may still be ``toy'' schemas. We take a systematic approach to identify and remove these in the analysis that follows.


\section{Schema Analysis: Characteristics and Comparisons}
\label{sec:analysis}

In this section, we analyze our GraphQL schema corpuses.
We discuss schema metrics, characterize the corpuses, and compare and contrast them.
Specifically, we
  analyze some general characteristics (\cref{sec:analysis-general}),
  identify naming conventions (\cref{sec:analysis-namingConventions}),
  estimate worst-case query response sizes (\cref{sec:analysis-responseSizes}),
  and measure the use of pagination, a common defense against pathological queries (\cref{sec:analysis-pagination}).

Because our purpose is to understand GraphQL practices in open-source and in industry, we extracted a subset of the GitHub corpus called the ``GitHub-large'' (GH-large) corpus that is comparable in complexity to the commercial corpus.
This distinction is useful for measurements that are dependent on the ``quality'' of the schema, e.g., worst-case response sizes and defenses, though for studies like trends in naming conventions we think it is appropriate to also consider the full GitHub corpus.
In future analyses, other measures of quality could be considered to segment the data, for example the number of stargazers of the associated GitHub repository.

\begin{figure}[h]
  \centering
  \includegraphics[width=\columnwidth]{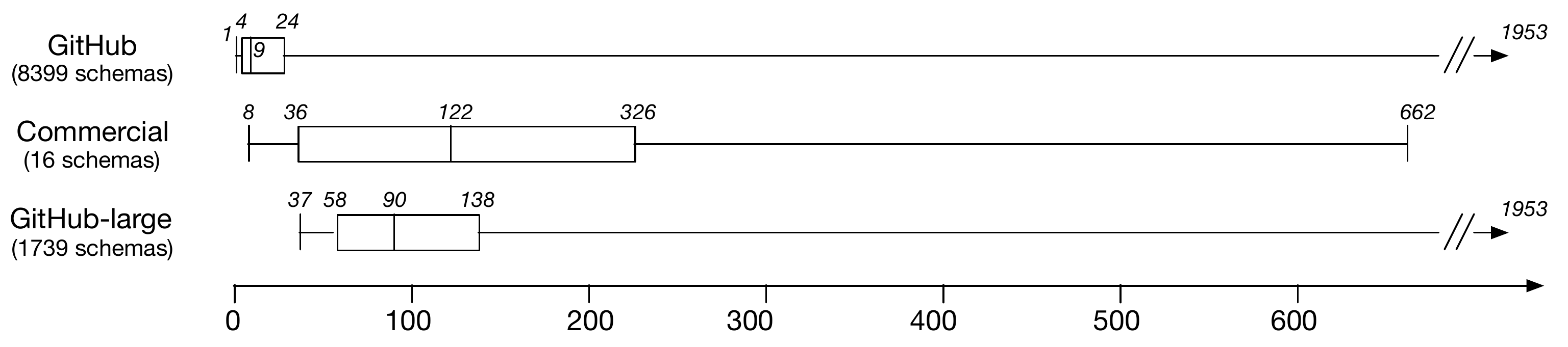}
  \vspace{-10pt}
  \caption{Distributions of schema complexity (number of definitions) in the GitHub, commercial, and GitHub-large schema corpuses. Whiskers show min and max values and the boxes show the quartiles.}
  \label{fig:data_distributions}
\end{figure}

We identified the GitHub-large corpus using a simple measure of schema complexity, namely its number of distinct definitions (for types, directives, operations etc.).
As shown in~\cref{fig:data_distributions}, the smallest commercial schema contains 8 definitions, while half of the GitHub corpus contains 9 or less definitions.
To avoid a bias toward these toy schemas and to accommodate the small sample size of the commercial corpus, we conservatively define a GitHub schema as \textit{large} if it is at least as complex as the first quartile of the commercial corpus (i.e., has more than $36$ definitions).
We include separate measurements on this GitHub-large corpus (\QualitySchemas schemas, \QualityCorpusPercentOfGithubCorpus of the GitHub corpus, \QualityMergedSchemas of which were recovered through merging).
The complexity distribution of the GitHub-large corpus is not perfectly aligned with the commercial corpus, but it is a better approximation than the GitHub corpus and allows more meaningful comparisons of open-source and industry GraphQL practices.


\subsection{Schema Characteristics} \label{sec:analysis-general}
First, we provide a reference for what GraphQL schemas look like in practice.
This snapshot can inform the design of GraphQL backends (e.g., appropriately sizing caches) as well as the development of the GraphQL specification (e.g., more and less popular features).
We parsed each schema using the graphql-js reference implementation~\cite{GraphQL-js} and analyzed the resulting AST.

\begin{table}[h]
  \footnotesize
  \centering
  \caption{Characteristics \& Features Used in Schema Corpuses.} 
  \label{table:schemas}
  \begin{tabular}{lccc}
       & \textbf{Commercial} (\CommercialSchemas) & \textbf{GitHub} (\Schemas) & \textbf{GH-large} (\QualitySchemas) \\
    \toprule
    Median object types (OTs) & \CommercialMedianOTs & \MedianOTs & \QualityMedianOTs \\
    Median input OTs          & \CommercialMedianInputOTs & \MedianInputOTs & \QualityMedianInputOTs \\
    Median fields in OTs               & \CommercialMedianFieldsOTs & \MedianFieldsOTs & \QualityMedianFieldsOTs \\
    Median fields in Input OTs         & \CommercialMedianFieldsInputOTs & \MedianFieldsInputOTs & \QualityMedianFieldsInputOTs \\
    \midrule
    Have interface types      & \CommercialWithInterfaces (\CommercialWithInterfacesPercent) & \WithInterfaces (\WithInterfacesPercent) & \QualityWithInterfaces (\QualityWithInterfacesPercent) \\
    Have union types          & \CommercialWithUnions (\CommercialWithUnionsPercent) & \WithUnions (\WithUnionsPercent) & \QualityWithUnions (\QualityWithUnionsPercent) \\
    Have custom directives    & \CommercialWithDirectives (\CommercialWithDirectivesPercent) & \WithDirectives (\WithDirectivesPercent) & \QualityWithDirectives (\QualityWithDirectivesPercent) \\
    Subscription support      & \CommercialWithSubscriptions (\CommercialWithSubscriptionsPercent) & \WithSubscriptions (\WithSubscriptionsPercent) & \QualityWithSubscriptions (\QualityWithSubscriptionsPercent) \\
    Mutation support                  & \CommercialWithMutations (\CommercialWithMutationsPercent) & \WithMutations (\WithMutationsPercent) & \QualityWithMutations (\QualityWithMutationsPercent) \\
  \bottomrule
  \end{tabular}
\end{table}
\vspace{-10pt}


\cref{table:schemas} shows clear differences among all three corpuses.
Not surprisingly, commercial and GitHub-large schemas are larger, containing more object and input object types.
The sizes of individual object and input object types, however, look similar in all corpuses.
In terms of feature use, commercial schemas apply interface types, union types, and custom directives most frequently, followed by GitHub-large schemas and then GitHub schemas.
Conversely, GitHub-large schemas have mutation and subscription\footnote{Subscriptions permit clients to register for continuous updates on data.} support most frequently, followed by GitHub schemas and then commercial schemas.

Analyzing multiple corpuses provides a fuller picture of GraphQL practices.
For example, suppose you were to propose changes to the GraphQL specification based solely on one of these corpuses, e.g. to identify little-used features as deprecation candidates.
Considering only commercial schemas, subscription support appears to be unpopular (none of the commercial schemas offer subscriptions), so subscriptions might be a deprecation candidate.
But the GitHub-large corpus tells a different story: subscriptions are offered in 64\% of the GitHub-large schemas.
Considering only the GitHub-large corpus instead, you might conclude that custom directives are a deprecation candidate (only 1.5\% of GitHub-large schemas use them), even though 12.5\% of the commercial corpuses use them.
In both cases a single-corpus analysis is misleading, showing the value of our multi-corpus analysis.

\begin{mdframed}[backgroundcolor=black!10]
  \textbf{Finding 1}:
  Commercial and GitHub-large schemas are generally larger than GitHub schemas. Reliance on different GraphQL features (e.g., unions, custom directives, subscription, mutation) varies widely by corpus.
\end{mdframed}


\subsection{Naming Conventions} \label{sec:analysis-namingConventions}
Naming conventions help developers understand new interfaces quickly and create interfaces that are easily understandable.
In this section we explore the prescribed and organic naming conventions that GraphQL schema authors follow, e.g. common ways to name types, fields, and directives.
\cref{table:conventions} summarizes our findings. We focus on the proportion of schemas that follow a convention \emph{consistently}, i.e., the schemas that use them in all possible cases.

\begin{table}[h]
  \footnotesize
  \centering
  \caption{The proportion of schemas that consistently adhere to prescribed (upper part) and organic (lower part) naming conventions. In rows marked with a $\dagger$ we report percentages from the subsets of schemas that use any enums, input object types, or mutations, respectively.}
  \label{table:conventions}
  \begin{tabular}{lccc}
                                & \textbf{Commercial} (\CommercialSchemas) & \textbf{GitHub} (\Schemas) & \textbf{GH-large} (\QualitySchemas) \\
    \toprule
    camelCase field names           & \CommercialCamelFieldSchemas         & \CamelFieldSchemas         & \QualityCamelFieldSchemas \\
    PascalCase type names           & \CommercialPascalTypeSchemas         & \PascalTypeSchemas         & \QualityPascalTypeSchemas \\
    PascalCase enum names $\dagger$ & \CommercialPascalEnumSchemas         & \PascalEnumSchemas         & \QualityPascalEnumSchemas \\
    ALL\_CAPS enum values $\dagger$ & \CommercialAllCapsEnumSchemas        & \AllCapsEnumSchemas        & \QualityAllCapsEnumSchemas \\
    \midrule
    \code{Input} postfix  $\dagger$ & \CommercialInputPostfixSchemas       & \InputPostfixSchemas       & \QualityInputPostfixSchemas \\
    Mutation field names  $\dagger$ & \CommercialMutationFieldNamesSchemas & \MutationFieldNamesSchemas & \QualityMutationFieldNamesSchemas \\
    snake\_case field names         & \CommercialSnakeFieldSchemas         & \SnakeFieldSchemas         & \QualitySnakeFieldSchemas \\
  \bottomrule
  \end{tabular}
\end{table}
\vspace{-10pt}

\subsubsection{Prescribed conventions.}
GraphQL experts have recommended a set of naming conventions through written guidelines~\cite{ApolloConventions:2019} as well as implicitly through the example schemas in the GraphQL documentation~\cite{GraphQLDocs}.
These prescribed conventions are: (1) Fields should be named in camelCase; (2) Types should be named in PascalCase; and (3) Enums should be named in PascalCase with (4) values in ALL\_CAPS.

We tested the prevalence of these conventions in real GraphQL schemas\footnote{For simplicity, we tested for camelCase and PascalCase names using only the first letter. A more sophisticated dictionary-based analysis is a possible extension.}.
As shown in \cref{table:conventions}, these prescribed conventions are far from universal.
The only prescribed convention that is frequently used in all three corpuses is (3) PascalCase enum names, exceeding 80\% of schemas in each corpuses and over 95\% in the GitHub and GitHub-large corpuses.
In contrast,
  (1) camelCase field names are only common in GitHub schemas,
  (2) PascalCase type names are common in GitHub and GitHub-large schemas and less so in commercial schemas,
  and (4) ALL\_CAPS enum values appear in more than half of commercial schemas, but are unusual in the GitHub and GitHub-large schemas.

\subsubsection{Organic conventions.}

Next we describe ``organic'' conventions\footnote{These conventions are ``organic'' in the sense that they are emerging naturally without apparent central direction. There could, however, be some hidden form of direction, e.g. many projects influenced by the same team or corporation.} that we observed in practice but which are not explicitly recommended in grey literature like the GraphQL specification or high-profile GraphQL tutorials.

\textbf{\code{Input} postfix for input object types.}
Schemas in our corpuses commonly follow the convention of ending the names of input object types with the word \code{Input}.
This convention is also followed in the examples in the official GraphQL documentation~\cite{GraphQLDocs}, but the general GraphQL naming convention recommendations do not remark on it~\cite{ApolloConventions:2019}.
In GraphQL, type names are unique, so the \code{Input} postfix is often used to associate object types with related input object types (e.g., the  object type \code{User} may be related to the input object type \code{UserInput}).

\textbf{Mutation field names.}
Developers commonly indicate the effect of the mutation by including it as part of the field name.
These names are similar to those used in other data contexts:
\code{update},
\code{delete},
\code{create},
\code{upsert},
and \code{add}.

\textbf{snake\_case field names.}
Of the non-camelCase field names in the GitHub corpus, \SnakeCasePercentOfNonCamelCase follow snake\_case (determined by the presence of an underscore: ``\code{\_}''), covering \SnakeField of all field names and used in \SomeSnakeCasePercent of all schemas in the GitHub corpus.
However, barely any schema across all corpuses uses this convention throughout.

In general, the observed organic conventions are much more common in GitHub and GitHub-large schemas than in commercial schemas.

\begin{mdframed}[backgroundcolor=black!10]
  \textbf{Finding 2}:
  GraphQL experts have recommended certain naming conventions. We found that PascalCase enum names are common in all three corpuses, and PascalCase type names are common in the GitHub and GitHub-large corpuses, but other recommendations appear less consistently. In addition, we observed the relatively common practice of \code{input} postfix and mutation field names in the GitHub and GitHub-large corpuses.  We recommend that commercial API providers improve the usability of their APIs by following both recommended and ``organic'' conventions.
\end{mdframed}

\subsection{Schema Topology and Worst-Case Response Sizes} \label{sec:analysis-responseSizes}

Queries resulting in huge responses may be computationally taxing, so practitioners point out the resulting challenge for providers to throttle such queries~\cite{Rinquin:2017,Stoiber:2018}.
The size of a response depends on three factors: the schema, the query, and the underlying data.
In this section, we analyze each schema in our corpuses for the worst-case response size it enables with pathological queries and data.

A GraphQL query names all the data that it retrieves (cf.~\cref{fig:example-query}).
Provided a schema has no field that returns a list of objects, the response size thus directly corresponds to the size of the query.
On the other hand, if a field can return a list of objects (e.g., \texttt{nodes} in \cref{fig:example-query}), nested sub-queries are applied to all the elements of the list (e.g., \texttt{name} in \cref{fig:example-query}).
Therefore, nested object lists can lead to an explosion in response size.

From the schema we can compute~$K$, the maximum number of nested object lists that can be achieved in a query.
For example, if \texttt{Query} contains a field \texttt{repos:[Repo]}, and \texttt{Repo} contains a field \texttt{members:[User]} then $K=2$.
Without access to the underlying data, we assume that the length of all the retrieved object lists is bounded by a known constant~$D$.\footnote{In practice, the size of retrieved object lists are often explicitly bounded by slicing arguments (e.g., \texttt{first: 2} in \cref{fig:example-query}). See also \cref{sec:analysis-pagination}.}

\textbf{Polynomial response.}
For a query of size~$n$, the worst-case response size is $O\!\left((n-K) \times D^K\right)$ --- at worst polynomial in the length~$D$ of the object lists.
The proof is by induction over the structure of the query.
As an illustration, consider the worst-case scenario of a query with maximum number of nested lists,~$K$.
Since the query must spend~$K$ fields to name the nested lists, each object at the deepest level can have at most $(n-K)$ fields and will be of size at most $(n-K)$.
Each level returns~$D$ nested objects, plus one field to name the list.
The size of each level~$k$ starting from the deepest one thus follows the relation: $s_k = D \times s_{k-1} + 1$ with $s_0 = (n-K)$.
The response size is given by the top level~$K$: $s_K = (n-K) \times D^K + \frac{D^K - 1}{D - 1}$, that is, $O\!\left((n-K) \times D^K\right)$.\footnote{In \cref{table:TypeGraphResponseComplexityAnalysis}, we use the slightly relaxed notion $O(n \times D^K)$.}

\textbf{Exponential response.}
If the schema includes a cycle containing list types (e.g., a type \texttt{User} contains a field \texttt{friends:[User]}), the maximum number of nested object lists is only bounded by the size of the query, i.e., $K < n$.\footnote{In GraphQL the first field is always \texttt{query}, and cannot be a list type.}
In that case the worst-case response size becomes $O(D^{n-1})$, that is, exponential in the size of the query.
Consider for example the following query that requests names of third degree friends (size $n=4$ and nesting $K=3$). 
If every user has at least ten friends, the size of the response is $1 + 10 \times (1 + 10 \times (1 + 10 \times 1)) = 1111$. 
\begin{lstgqlsmall}
query { friends(first: 10) { friends(first: 10) { friends(first: 10) { name } } } }
\end{lstgqlsmall}

\begin{table}[h!]
\footnotesize
\centering
\caption{Worst-case response size based on type graph analysis, where~$n$ denotes the query size, and~$D$ the maximum length of the retrieved lists.}
\label{table:TypeGraphResponseComplexityAnalysis}
\begin{tabular}{llccc}
\multicolumn{2}{l}{\textbf{Worst-case response}} & \textbf{Commercial} (\CommercialSchemas) & \textbf{GitHub} (\Schemas) & \textbf{GH-large} (\QualitySchemas) \\ \toprule
Exponential & $O(D^{n-1})$          &  \CommercialWorstExponential (\CommercialWorstExponentialPercent) & \WorstExponential (\WorstExponentialPercent)     &  \QualityWorstExponential (\QualityWorstExponentialPercent) \\ \midrule
Polynomial & $O(n \times D^6)$             &  \CommercialWorstSix (\CommercialWorstSixPercent)      &  \WorstSix (\WorstSixPercent) &  \QualityWorstSix (\QualityWorstSixPercent) \\
Polynomial & $O(n \times D^5)$             &  \CommercialWorstFive (\CommercialWorstFivePercent)    &  \WorstFive (\WorstFivePercent) &  \QualityWorstFive (\QualityWorstFivePercent) \\
Polynomial & $O(n \times D^4)$             &  \CommercialWorstFour (\CommercialWorstFourPercent)    &  \WorstFour (\WorstFourPercent) &  \QualityWorstFour (\QualityWorstFourPercent) \\
Polynomial & $O(n \times D^3)$             &  \CommercialWorstThree (\CommercialWorstThreePercent)  &  \WorstThree (\WorstThreePercent) &  \QualityWorstThree (\QualityWorstThreePercent) \\
Quadratic & $O(n \times D^2)$              &  \CommercialWorstTwo (\CommercialWorstTwoPercent)      &  \WorstTwo (\WorstTwoPercent) &  \QualityWorstTwo (\QualityWorstTwoPercent) \\ \midrule
Linear & $O(n \times D)$                   &  \CommercialWorstOne (\CommercialWorstOnePercent)      &  \WorstOne (\WorstOnePercent) &  \QualityWorstOne (\QualityWorstOnePercent) \\
Linear & $O(n)$                            &  \CommercialWorstZero (\CommercialWorstZeroPercent)    &  \WorstZero (\WorstZeroPercent) &  \QualityWorstZero (\QualityWorstZeroPercent) \\
\bottomrule
\end{tabular}
\end{table}

\textbf{Results.}
We implemented an analysis for schema topographical connectedness based on the conditions for exponential and polynomial responses sizes outlined above, and applied it to our schema corpuses.
As shown in~\cref{table:TypeGraphResponseComplexityAnalysis}, the majority of commercial (\CommercialSuperLinear), GitHub (\SuperLinear), and GitHub-large (\QualitySuperLinear) schemas have super-linear worst-case response sizes.
This finding is of course not altogether surprising, as the key to super-linear response sizes is a particular and intuitive relational schema structure, and the purpose of GraphQL is to permit schema providers to describe relationships between types.
However, the implication is that GraphQL providers and middleware services should plan to gauge the cost of each query by estimated cost or response size, or otherwise limit queries.

\begin{mdframed}[backgroundcolor=black!10]
  \textbf{Finding 3}:
  The majority of commercial, GitHub, and GitHub-large schemas have super-linear worst-case response sizes, and in the commercial and GitHub-large corpuses, they are mostly exponential.
  Providers need to consider throttling requests to their APIs to avoid the negative consequences of expensive queries, whether malicious or inadvertent.
\end{mdframed}


\subsection{Delimiting Worst-Case Response Sizes through Pagination}\label{sec:analysis-pagination}


Queries with super-linear response sizes can become security threats, overloading APIs or even leading to denial-of-service.
For commercial GraphQL providers, exponential response sizes pose a potential security risk (denial of service).
Even polynomial response sizes might be concerning --- e.g., consider the cost of returning the (very large) cross product of all GitHub repositories and users.

The official GraphQL documentation recommends that schema developers use one of two \emph{pagination} techniques to bound response sizes~\cite{GraphQLDocs-Pagination}:
\textbf{Slicing} refers to the use of numeric arguments to index a subset of the full response set.
The \textbf{connections pattern} introduces a layer of indirection to enable more complex pagination.
The addition of \textit{Edge} and \textit{Connection} types allows schema developers to indicate additional relationships between types, and to paginate through a concurrently updated list (cf. schema described in Section~\ref{sec:background}).



\textbf{Analysis.} We used heuristics relying on names of fields and types to identify the use of pagination patterns within schemas.

For slicing, we identify fields that return object lists and accept numeric \textit{slicing arguments} of scalar type \code{Int}.
In our corpuses these arguments are commonly named \code{first}, \code{last}, and \code{limit}, or \code{size}.
We use the presence of arguments with these names as an indication that slicing is in use.
We differentiate schemas that use such arguments for slicing consistently, for some fields, or not at all.

For the connections pattern, we check schemas for types whose names end in \code{Connection} or \code{Edge} as proposed in the official GraphQL docs~\cite{GraphQLDocs-Pagination}.
We again check for the use of slicing arguments on fields that return connections.

\begin{table}
  \footnotesize
  \centering
  \caption{Use of Slicing Arguments and Connections Pattern.}
  \label{table:pagination}
  \begin{tabular}{lccc}
                                      & \textbf{Comm.} (\CommercialSchemas)                                                           & \textbf{GitHub} (\Schemas)                 & \textbf{GH-large} (\QualitySchemas) \\
  \toprule
  Have fields returning object lists  & \CommercialNumSchemasLists (\CommercialNumSchemasListsPercent)                       & \NumSchemasLists (\NumSchemasListsPercent)               & \QualityNumSchemasLists (\QualityNumSchemasListsPercent) \\
  \ ...with no slicing arguments      & \CommercialNumSchemasNoSlicing (\CommercialNumSchemasNoSlicingPercent)               & \NumSchemasNoSlicing (\NumSchemasNoSlicingPercent)       & \QualityNumSchemasNoSlicing (\QualityNumSchemasNoSlicingPercent) \\
  \ ...with slicing args. sometimes   & \CommercialNumSchemasMixedSlicing (\CommercialNumSchemasMixedSlicingPercent)         & \NumSchemasMixedSlicing (\NumSchemasMixedSlicingPercent) & \QualityNumSchemasMixedSlicing (\QualityNumSchemasMixedSlicingPercent) \\
  \ ...with slicing args. throughout  & \CommercialNumSchemasPureSlicing (\CommercialNumSchemasPureSlicingPercent)  & \NumSchemasPureSlicing (\NumSchemasPureSlicingPercent)   & \QualityNumSchemasPureSlicing (\QualityNumSchemasPureSlicingPercent) \\
  \midrule
  Have types with names matching \\\code{/Edge\$/} and \code{/Connection\$/} & \CommercialNumSchemasConnections (\CommercialNumSchemasConnectionsPercent) & \NumSchemasConnections (\NumSchemasConnectionsPercent) & \QualityNumSchemasConnections (\QualityNumSchemasConnectionsPercent) \\
  \ ...with no slicing arguments     & \CommercialNumSchemasConnectionsNoSlicing (\CommercialNumSchemasConnectionsNoSlicingPercent) & \NumSchemasConnectionsNoSlicing (\NumSchemasConnectionsNoSlicingPercent) & \QualityNumSchemasConnectionsNoSlicing (\QualityNumSchemasConnectionsNoSlicingPercent) \\
  \ ...with slicing args. sometimes  & \CommercialNumSchemasConnectionsMixedSlicing (\CommercialNumSchemasConnectionsMixedSlicingPercent) & \NumSchemasConnectionsMixedSlicing (\NumSchemasConnectionsMixedSlicingPercent) & \QualityNumSchemasConnectionsMixedSlicing (\QualityNumSchemasConnectionsMixedSlicingPercent) \\
  \ ...with slicing args. throughout & \CommercialNumSchemasConnectionsPureSlicing (\CommercialNumSchemasConnectionsPureSlicingPercent) & \NumSchemasConnectionsPureSlicing (\NumSchemasConnectionsPureSlicingPercent) & \QualityNumSchemasConnectionsPureSlicing (\QualityNumSchemasConnectionsPureSlicingPercent) \\
  \bottomrule
  \end{tabular}
\end{table}

\textbf{Results.} Using our heuristics, \cref{table:pagination} summarizes the use of the pagination patterns in our corpuses.
In no corpus are these pagination patterns used consistently, strengthening the threat of the worst-case responses discussed in~\cref{sec:analysis-responseSizes}.
For the schemas that do use pagination patterns, the commercial and GitHub-large schemas tend to use the more complex yet flexible connections pattern, while slicing alone is used inconsistently across all schemas.


\begin{mdframed}[backgroundcolor=black!10]
  \textbf{Finding 4}:
  No corpus consistently uses pagination patterns, raising the specter of worst-case response sizes.
  When pagination patterns are used, commercial and GitHub-large schemas tend to use the connections pattern, while slicing is not used consistently.
  Our worst-case findings from~\cref{sec:analysis-responseSizes} urge the wider adoption of pagination.
\end{mdframed}

\section{Related Work}
\label{sec:related_work}

Our work is most closely related to that of Kim et al., who also collected and analyzed GraphQL schemas~\cite{Kim:2019}.
They analyzed 2,081 unique schemas mined from open-source repositories on GitHub.
Our works are complementary. We use different mining techniques and conduct different analyses.
For \textbf{mining}, to identify GraphQL schemas on GitHub, both works queried the GitHub API for filenames with GraphQL-themed substrings.
We additionally proposed a novel schema stitching technique to repair incomplete schemas, which permitted us to recover thousands of schemas that their methodology would discard (\cref{section:corpuses-github}).
In \textbf{analysis}, we compared multiple corpuses, while they focused solely on schemas obtained from GitHub and did not distinguish between the larger and smaller schemas therein.
Where our analyses overlap, our findings agree:
 in our GitHub schema corpus we report
   similar proportions of schemas using mutations (we: \WithMutationsPercent, they: $70\%$)
   and
   subscriptions (we: \WithSubscriptionsPercent, they: $20\%$).
Similarly, in our GitHub corpus we found a similar proportion of schemas with type cycles (we: \WorstExponentialPercent, they: $39.7\%$).
Our analyses of naming conventions, worst-case response sizes, and pagination are novel.

Our worst-case response size analysis (\cref{sec:analysis-responseSizes}) benefits from the work of Hartig and P\'erez.
They complemented the GraphQL specification~\cite{GraphQL-Spec} with a formal description for key parts of GraphQL~\cite{Hartig2017GQLWorkshop,Hartig2018GQLSemantics}.
They also proved the existence of GraphQL schema-data (graph) combinations on which a query will have exponential-sized results (cf.~\cite[Propositions~5.2 and 5.3]{Hartig2018GQLSemantics}) and gave an upper bound for the response size (cf.~\cite[Theorem~5.4]{Hartig2018GQLSemantics}).
In comparison, our analysis in~\cref{sec:analysis-responseSizes}
explicitly identifies object lists as the cause of the response size explosion, and we use this observation to provide a tighter upper bound.

The remaining academic literature on GraphQL focuses on the challenges of creating a GraphQL API.
Several research teams have described their experiences exposing a GraphQL API or migrating existing APIs to GraphQL~\cite{Vogel2017,Vazquez-Ingelmo2017,Brito:2019}.
Others have described automatic techniques for migration~\cite{Wittern:2018} and testing~\cite{Vargas2018}.

Our work is similar in spirit to studies of REST(-like) APIs, which have focused on API design best practices~\cite{Palma:2015,Petrillo:2016} or assessed API business models~\cite{Gamez:2017}.
Because of the paradigmatic differences between GraphQL and REST (single endpoint, typed schema, queries formed by clients, etc.), this work complements existing ones.

\section{Threats to Validity} \label{sec:threats}

\noindent \textbf{Construct validity.}
In~\cref{sec:analysis-responseSizes} we assume that response size is the primary measure of query cost.
We leave to future work a more fine-grained analysis dependent on backend implementation details (e.g. resolver function costs).
\vspace{0.2em}

\noindent \textbf{Internal validity.}
Our name-based analyses depend on heuristics which could be inaccurate,
although they are grounded in the grey literature where possible.
\vspace{0.2em}

\noindent \textbf{External validity.}
Our corpuses may not be representative of the true state of GraphQL schemas in practice, affecting the generalizability of our results. 
The commercial corpus contains the \CommercialSchemas public commercial GraphQL APIs we could identify, well short of the 100+ companies that use GraphQL (presumably internally)~\cite{GraphQLUsers}.
We restricted the open-source corpus to statically defined schemas stored in GitHub.
By analyzing the ``GitHub large'' schemas separately, we provide a better understanding of both (1) methodologically, the risks of treating all GitHub schemas alike, and (2) scientifically, the properties of larger schemas.

\section{Conclusions} \label{sec:conclusion}

GraphQL is an increasingly important technology.
We provide an empirical assessment of the current state of GraphQL through our rich corpuses, novel schema reconstruction methodology, and novel analyses.
Our characterization of naming conventions 
can help developers adopt community standards to improve API usability.
We have confirmed the fears of practitioners and warnings of researchers about the risk of denial of service against GraphQL APIs: most commercial and large open-source GraphQL APIs may be susceptible to queries with 
exponential-sized responses.
We report that 
many schemas do not follow best practices and thus incompletely defend against such queries.

Our work motivates many avenues for future research, such as: refactoring tools to support naming conventions, coupled schema-query analyses to estimate response sizes in middleware (e.g. rate limiting), and data-driven backend design.

\section{Acknowledgments} \label{sec:ack}

We are grateful to A. Tantawi, A. Kazerouni, and B. Pirelli for their feedback on the manuscript, and to O. Hartig for a helpful discussion.

%
%
\bibliographystyle{splncs04}
\bibliography{bibliography}

\begin{thebibliography}{10}
\providecommand{\url}[1]{\texttt{#1}}
\providecommand{\urlprefix}{URL }
\providecommand{\doi}[1]{https://doi.org/#1}

\bibitem{APIsGuru-GraphQLAPIs}
{APIs-guru/graphql-apis: A collective list of public GraphQL APIs}, {Available
  at \url{https://github.com/APIs-guru/graphql-apis}}

\bibitem{GraphiQL}
{GraphiQL: An in-browser IDE for exploring GraphQL}, {Available at
  \url{https://github.com/graphql/graphiql}}

\bibitem{GraphQL-Faker}
{GraphQL Faker}, {Available at
  \url{https://github.com/APIs-guru/graphql-faker}}

\bibitem{ApolloConventions:2019}
{GraphQL Style conventions}, {Available at
  \url{https://www.apollographql.com/docs/apollo-server/essentials/schema.html\#style}}

\bibitem{GraphQLDocs}
{Introduction to GraphQL}, {Available at \url{https://graphql.org/learn/}}

\bibitem{GraphQL-js}
{JavaScript reference implementation for GraphQL}, {Available at
  \url{https://github.com/graphql/graphql-js}}

\bibitem{GraphQLDocs-Pagination}
{Pagination}, {Available at \url{http://graphql.github.io/learn/pagination/}}

\bibitem{GraphQLSchemaLanguage:2018}
{Schemas and Types}, {Available at \url{https://graphql.org/learn/schema}}

\bibitem{GraphQLUsers}
{Who’s using GraphQL?}, {Available at \url{http://graphql.org/users}}

\bibitem{YelpGraphQL:2017}
{Introducing Yelp's Local Graph} (2017), {Available at
  \url{https://engineeringblog.yelp.com/2017/05/introducing-yelps-local-graph.html}}

\bibitem{NYTimesGraphQL:2017}
{React, Relay and GraphQL: Under the Hood of The Times Website Redesign}
  (2017), {Available at
  \url{https://open.nytimes.com/react-relay-and-graphql-under-the-hood-of-the-times-website-redesign-22fb62ea9764}}

\bibitem{GitHubGraphQL:2019}
{GitHub GraphQL API v4} (2019), {Available at
  \url{https://developer.github.com/v4/}}

\bibitem{ShopifyGraphQL:2019}
{GraphQL and Shopify} (2019), {Available at
  \url{https://help.shopify.com/en/api/custom-storefronts/storefront-api/graphql/}}

\bibitem{GraphQLDocs-Executable}
{GraphQL Current Working Draft: Schema} (2019), {Available at
  \url{https://facebook.github.io/graphql/draft/#sec-Executable-Definitions}}

\bibitem{GraphQLDocs-Schema}
{GraphQL Current Working Draft: Schema} (2019), {Available at
  \url{https://facebook.github.io/graphql/draft/#sec-Schema}}

\bibitem{Brito:2019}
Brito, G., Mombach, T., Valente, M.T.: {Migrating to GraphQL: A Practical
  Assessment}. In: 2019 IEEE 26th International Conference on Software
  Analysis, Evolution and Reengineering (SANER). pp. 140--150. IEEE (2019)

\bibitem{GraphQL-Spec}
{\relax Facebook Inc.}: {GraphQL. Working Draft} (June 2018), {Available at
  \url{https://facebook.github.io/graphql/}}

\bibitem{Gamez:2017}
Gamez-Diaz, A., Fernandez, P., Ruiz-Cortes, A.: {An Analysis of RESTful APIs
  Offerings in the Industry}. In: International Conference on Service-Oriented
  Computing. pp. 589--604. Springer (2017)

\bibitem{Hartig2017GQLWorkshop}
Hartig, O., P{\'{e}}rez, J.: {An initial analysis of Facebook's GraphQL
  language}. In: CEUR Workshop Proceedings (2017)

\bibitem{Hartig2018GQLSemantics}
Hartig, O., P{\'{e}}rez, J.: {Semantics and Complexity of GraphQL}. In:
  Conference on World Wide Web (WWW) (2018)

\bibitem{Kim:2019}
Kim, Y.W., Consens, M.P., Hartig, O.: {An Empirical Analysis of GraphQL API
  Schemas in Open Code Repositories and Package Registries}. In: Proceedings of
  the 13th Alberto Mendelzon International Workshop on Foundations of Data
  Management (AMW) (Jun 2019)

\bibitem{Palma:2015}
Palma, F., Gonzalez-Huerta, J., Moha, N., Gu{\'e}h{\'e}neuc, Y.G., Tremblay,
  G.: {Are RESTful APIs Well-Designed? Detection of their Linguistic (Anti)
  Patterns}. In: Int. Conf. on Service-Oriented Computing. pp. 171--187.
  Springer (2015)

\bibitem{Petrillo:2016}
Petrillo, F., Merle, P., Moha, N., Gu{\'e}h{\'e}neuc, Y.G.: {Are REST APIs for
  cloud computing well-designed? An exploratory study}. In: International
  Conference on Service-Oriented Computing. pp. 157--170. Springer (2016)

\bibitem{Rinquin:2017}
Rinquin, A.: {Avoiding n+1 requests in GraphQL, including within subscriptions}

\bibitem{Stoiber:2018}
Stoiber, M.: {Securing Your GraphQL API from Malicious Queries}

\bibitem{Vargas2018}
Vargas, D.M., Mayor, U., Sim, D.S., Blanco, A.F., Pablo, J., Alcocer, S.,
  Torres, M.M., Bergel, A.: {Deviation Testing: A Test Case Generation
  Technique for GraphQL APIs} (2018)

\bibitem{Vazquez-Ingelmo2017}
V{\'{a}}zquez-Ingelmo, A., Cruz-Benito, J., Garc{\'{i}}a-Pe{\~{n}}alvo, F.J.:
  {Improving the OEEU's data-driven technological ecosystem's interoperability
  with GraphQL}. In: Proceedings of the 5th International Conference on
  Technological Ecosystems for Enhancing Multiculturality - TEEM 2017. pp.~1--8
  (2017)

\bibitem{Vogel2017}
Vogel, M., Weber, S., Zirpins, C.: {Experiences on Migrating RESTful Web
  Services to GraphQL}. In: ICSOC Workshops. Springer International Publishing
  (2017)

\bibitem{Wittern:2019b}
Wittern, E., Cha, A., Davis, J.C., Baudart, G., Mandel, L.: {GraphQL Schema
  Collector}. \doi{10.5281/zenodo.3352421}, accessible at
  \url{https://github.com/ErikWittern/graphql-schema-collector}

\bibitem{Wittern:2019}
Wittern, E., Cha, A., Davis, J.C., Baudart, G., Mandel, L.: {GraphQL Schemas}.
  \doi{10.5281/zenodo.3352419}, accessible at
  \url{https://github.com/ErikWittern/graphql-schemas}

\bibitem{Wittern:2018}
Wittern, E., Cha, A., Laredo, J.A.: {Generating GraphQL-Wrappers for REST
  (-like) APIs}. In: International Conference on Web Engineering. pp. 65--83.
  Springer (2018)

\end{thebibliography}
%




\end{document}